**Refining the conference experience for junior scientists in the wake of climate change**

Ruth Johnson[1], Andrada Fiscutean[2], Serghei Mangul[3]

[1] Department of Computer Science, University of California, Los Angeles, 404 Westwood Plaza, Los Angeles, CA 90024, USA

[2] Independent technology journalist

[3] Department of Clinical Pharmacy, School of Pharmacy, University of Southern California, 1985 Zonal Avenue Los Angeles, CA 90089, USA

**Abstract**

With the ever-increasing carbon footprint associated with conferences, scientists can learn to refine their conference experiences when they do need to travel. We offer insight on how to optimize the conference experience through attending speaker sessions, giving presentations, and networking.

**Main text**

Conference travel is a key component of the academic job—a catalyst to the development of new ideas and collaborations that foster innovation and bring together the research community[1]. The in-person interactions—sometimes planned and many times spurious—are vital to the sharing of ideas across disciplines and often generate career-crucial networking more effectively than interacting over video conferencing. However, it is clear that academic travel contributes to human-induced climate change with thousands of academics jetting hundreds of thousands of miles around the globe each year[2–6]. Both current academics and upcoming trainees are realizing the role that this routine plays in our carbon footprint and are embracing the adoption of the 3R's: "replace, reduce, refine"[7]. Senior researchers have already cultivated a level of authority and reputation that allows them to abstain from conference travel—and it would likely be a big statement if they did. But early-stage researchers might not have this option. For junior scholars, in-person networking can translate into opportunities to showcase their work, better their career prospects, and gain a more established role in the research community. Nascent scientists do not have the luxury to replace and reduce, but we can focus on refining their experiences when they do travel.

Optimizing time spent when traveling is difficult because there is a lack of resources for navigating conferences specifically written for junior scientists. Trainees are constantly mentored on how to do research, but they are rarely told how to communicate these ideas or to network with other individuals interested in this research. Based off of our own experiences, we offer advice about how to maximize time and travel resources that are both accessible and advantageous for junior scientists.

**Attending sessions**

At conferences, effectiveness is oftentimes overlooked in the pursuit of efficiency. There is the misconception that researchers must attend every session or jump between parallel sessions in order to achieve maximum efficiency at a meeting. Attending every single talk can deplete the attendee's attention span that may be necessary for important talks later in the day. We find it helpful to carefully plan, ahead of time, which sessions should be attended while explicitly considering the costs and benefits of moving in between sessions and retaining necessary energy levels for the remainder of the day. We suggest factoring into the daily schedule sufficient time to conserve, restore, and regroup energy throughout the day—even if that means sitting out a session that seems relevant to one's own research.

We gain the most out of presentations by being active participants; therefore, it is important to be mindful of energy expenditure when attending sessions. Framing the presented research with how it fits with current research projects helps an attendee actively process new information. Asking questions is another effective way to engage in the presentation and create a discussion around relevant points. An attendee may find it useful to write down in advance questions for each talk, even if they will never be asked, because this helps one to think beyond what is presented on the slides.

**Giving presentations**

Showcasing their work can boost a junior researcher's career, but the overwhelming fear of presenting can make many junior scientists avoid delivering presentations at conferences.

Although a presentation largely consists of a visual component, it is necessary to adequately prepare the accompanying oral presentation so that the ideas can be clearly communicated, and the feedback can be maximized. Some of the best presentations sound natural and unrehearsed, but this is just an illusion—the speakers probably have gained years of practice. Junior researchers can rapidly improve their oral presentation skills and reduce presentation-related anxiety by holding practice sessions for colleagues and in front of a variety of groups, which allows questions from various audience perspectives. In addition, if a platform talk is only 10 or 15 minutes long--a length that does not allow much room for nervous pauses or superfluous phrases--do not hesitate to memorize key phrases or sentences. Common worries of junior presenters are about forgetting points or freezing during a presentation. One solution is to add the planned text into the speaker notes; it can be calming knowing that there is a backup plan on stage.

It is important to note that presentations are an integral part of receiving feedback in academia. Even though the question and answer portion of a talk is only a fraction of the overall time, these few minutes require as much preparation as does the main speaking portion. Write down any questions you believe someone may ask. Since you know all the intricate details of your work, you are the one who can come up with the hardest questions. Before presenting, review the key papers related to your presentation so they are fresh on your mind in case an audience member asks about them. Additionally, print out a copy of your slides beforehand and bring them with you; then directly after your talk, write down any questions, comments and feedback from the audience members. Finally, remember that no matter how much one prepares, a presentation will not be perfect. Each presentation is an opportunity to learn, so it is unproductive to be too critical of one's self.

**Utilizing coffee breaks and social events**

The time spent with colleagues outside of sessions is an essential component of conferences that is not available through digital communication platforms, so it is especially important to be aware of how this time can be used by a junior researcher. Coffee breaks between sessions and receptions in the evenings provide an opportunity to talk with speakers from earlier in the day or meet new colleagues. If a session is especially relevant to your research, then you can reserve that next coffee break for following up with some of the speakers. Additionally, larger conferences can bring together people from all over the world and from different disciplines. If there is a particular researcher from another university or research area you want to speak with, then reach out the week beforehand to set a time when you both can meet at the conference.

Conservation of the attendee's own physical energy is important, but we should also be conscious of how to properly allocate mental energy. Just as attentively listening during sessions uses a lot of energy, so does talking with others after the sessions. Networking brings up social anxiety, be it mild or severe, for many people. It is important to plan for these social settings and take the time to re-energize however you feel is needed. In particular, it can be useful to plan out how to balance one's social capacity over the week. For example, one can stay in the weekend before a conference to refuel or, if there is an evening event, one can take time alone during the coffee breaks. And if one struggles with initiating conversations at networking events, it is important to remember that everyone is attending because they are interested in the specific topic of the meeting. An easy conversation starter is simply asking about the research they do.

**Last remarks**

There are many reasons that make it difficult if not impossible for trainees and professors to attend conferences. Most of the ones listed above have to do with the inner challenges one might face, but there is also the difficulty of securing funding to attend, traveling with children or finding childcare, travel visa challenges, and taking care of loved ones at home, among others. If you are not able to travel for conferences, one could try organizing a local meeting within the university or surrounding region. Research conferences blur the line of one's professional goals and personal circumstances, presenting a unique challenge, but we hope that these points above will at least be useful to many who might garner similar thoughts, which most people are more than eager to share.